\documentstyle[preprint,aps,prc,epsf]{revtex}
\tightenlines
\begin{document}
\draft
\title{Hyperon-nucleon 
scattering and hyperon masses in the
nuclear medium}
\author{C.~L.~Korpa}
\address{Department of Theoretical Physics,
University of Pecs, Ifjusag u.\ 6, H-7624 Pecs, Hungary}
\author{A.~E.~L.~Dieperink and R.~G.~E.~Timmermans}
\address{Kernfysisch Versneller Instituut, Zernikelaan 25, NL-9747AA
Groningen, The Netherlands}
\date{\today}
\maketitle

\begin{abstract}
We analyze low-energy hyperon-nucleon scattering using an effective
field theory in next-to-leading order. 
By fitting experimental  
cross sections for laboratory hyperon momenta below 200 MeV$/c$
and using information from the hypertriton we determine twelve 
contact-interaction coefficients. 
Based on these we discuss the low-density expansion of hyperon 
mass shifts in the nuclear medium.

\end{abstract}


\section{Introduction}
The binding energy of hyperons in nuclear matter plays an important role
in hypernuclei and in the equation of state of neutron stars.
The most commonly used approach is that one starts from a two-body hyperon-nucleon 
potential based upon one-boson exchange (OBE) and SU(3)-flavor symmetry and then
computes the binding of hyperons in matter in the Brueckner approximation.
It has been shown that  uncertainties in the two-body input may lead to
large differences in the resulting hyperon potentials and in particular
in the binding of the sigma hyperons \cite{Dabrowski}.
Therefore it is appropriate to see whether other approaches, e.g.\
a low-density expansion based on vacuum scattering amplitudes can 
provide additional insight.

Alternatively, Savage and Wise \cite{Savage96} have analyzed hyperon mass shifts
in nuclear matter using chiral perturbation theory  by expanding in the Fermi
momentum. The interactions were determined by a chiral Lagrangian
with $\mbox{SU(3)}\times\mbox{SU(3)}$ symmetry; the hyperon mass shifts
could be expressed in terms of  six  coefficients related to the strength
of the four-baryon contact interactions and the pseudoscalar-meson couplings to 
baryons and were computed at the one-loop level. However, the unknown values of
these six parameters prevented them from obtaining numerical results
for the energy shift at zero momentum.

More recently, 
applications of effective field theory (EFT) in nuclear physics have 
received a renewed interest (for a review see 
\cite{Kolck99,Beane00}). So far, most
investigations have been devoted to the two-nucleon system 
\cite{Kaplan98,Gegelia99,Mehen99}, which is characterized by 
anomalously large scattering lengths. The existence of  
quasi-bound states introduces an additional scale in the
problem, complicating the power counting required in EFT. A remedy was recently 
proposed \cite{Kaplan98} in the form of a selective resummation. 
The  
leading-order amplitude of 
this approach reproduces the effective-range expansion with a
vanishing 
range parameter \cite{Kolck99a}. Inclusion of the next-to-leading order terms
produces a non-zero effective range. 

It is tempting to study whether the approach of Ref.\ \cite{Kaplan98} 
can be applied to the
hyperon-nucleon sector which involves coupled channels, i.e.\ the scattering
length and the effective range are represented by matrices. Since our region of
application (center-of-mass momenta below 100 MeV/$c$) is below the pion cut, 
and in view of the problems encountered with perturbative pions \cite{Fleming}, 
we use an approach without explicit mesons. 
Thus the two-baryon interaction to leading order involves 6 constant contact terms, since
SU(3)$\times$SU(3) can be decomposed into the sum of 6 irreducible 
representations of SU(3).
For the s-waves we consider, the next-to-leading (NLO) order introduces  
6 additional SU(3)-symmetry respecting coefficients
for the terms of order $p^2$. The absence of experimental information on
$\Xi N$ scattering means that the low-energy hyperon-nucleon data in the present 
approach depend only on 10 (out of these 12) parameters.  

As discussed in the following sections, breaking of SU(3) symmetry by
meson masses has significant consequences. We model this
symmetry breaking by incorporating terms of order $p^2$ coming from
one-pion exchange. The two coefficients of these terms we treat as free
parameters, resulting in a total of 12 parameters.


In Section II we formulate the coupled-channels formalism, based on 
effective field theory, for low-energy hyperon-nucleon scattering.
In Section III we present results of the fit to $YN$ cross sections and 
$\Lambda N$ 
scattering lengths. Some information on the latter is provided by the 
calculation of the hypertriton binding energy \cite{Miyagawa95}.
Results for the hyperon mass shifts in low-density nuclear medium
and their comparison to results obtained by other methods are
presented in Section IV.

\section{Hyperon-nucleon scattering}
We consider hyperon-nucleon scattering at low energy and use
the approach of effective field theory.
In general the tree-level amplitude in next-to-leading order is written as
\begin{equation}
{\cal{A}}_0=-C_0-C_2 p^2,\label{treeamplitude}
\end{equation}
where $C_0$ and $C_2$ denote matrices whose elements are the
relevant coefficients in the interaction Lagrangian. 

The Kaplan-Savage-Wise (KSW) 
resummation \cite{Kaplan98} gives, in leading order, a scattering amplitude $\cal A$ given
by
\begin{equation}
{\cal A}^{-1} = {\cal A}_0^{-1}-\frac{M_r}{2\pi}(\mu+ip),
\end{equation}
with $M_r$ the reduced baryon mass (for which we take a common value
in all channels) and $\mu$ the (arbitrary) subtraction
point.
Using the relation between the $K$-matrix  and the full amplitude $\cal A$, viz.
\begin{equation}
pK^{-1}=\frac{2\pi}{M_r}{\cal A}^{-1}+ip
\end{equation}
we obtain for the KSW resummation 
\begin{equation}
pK^{-1}=-\mu-\frac{2\pi}{M_r}(C_0+C_2 p^2)^{-1}.\label{pkmin}
\end{equation}
By expanding the term $(C_0+C_2p^2)^{-1}$ in powers of $p^2$
we recover the effective-range expansion 
\begin{equation}
pK^{-1}=-\frac{1}{a}+\frac{1}{2}rp^2+\cdots,
\end{equation}
where the scattering-length
matrix $a$ and effective-range matrix $r$ are given by
\begin{eqnarray}
a^{-1} &=& \mu+\frac{2\pi}{M_r}C_0^{-1},\\
r &=& \frac{4\pi}{M_r}C_0^{-1}C_2C_0^{-1}.
\end{eqnarray}
To accomodate possibly large values of the scattering length in the momentum
expansion of the amplitude one should keep all powers of $ap$. This
leads to the following expression for the amplitude in NLO:
\begin{equation}
{\cal A}=-\frac{2\pi}{M_r}\left( \mu+ip+\frac{2\pi}{M_r}C_0^{-1}\right)^{-1}
\left[ 1+\frac{2\pi p^2}{M_r}C_0^{-1}C_2 C_0^{-1}
\left( \mu+ip+\frac{2\pi}{M_r}C_0^{-1}\right)^{-1}
\right].
\end{equation}
This is a generalization of the single-channel expression, cf.\ Eq.\ 
(2.18) in Ref.\ \cite{Kaplan98}, to coupled channels.

Next we turn to the relationship between elements of the 
tree-level amplitude matrix $C_0$, corresponding to the two-baryon scattering 
amplitudes of interest. For the spin-1/2 baryon octet
it is convenient to express the 6 real parameters determining the constant 
four-baryon contact interactions  
through coefficients corresponding to the 6 irreducible representations 
of the direct product SU(3)$\times$SU(3), which is 
appropriate for baryon-baryon scattering.
We write the coefficients of the terms in the Lagrangian for 
B$_1$B$_2 \rightarrow $B$_3$B$_4$ 
in the form 
$-b_S({\rm{B}}_1{\rm{B}}_2\rightarrow {\rm{B}}_3{\rm{B}}_4)/f^2$,
where $f\approx 132$ MeV can be identified with the pion decay 
constant and $S=0$ or 1 denotes the total spin. Introducing the
notation $s_0,s_1,s_2,t_1,t_2,t_3$ for the coefficients of the
representations [1], [27], [8$_s$], [$\overline{10}$], 
[10], [8$_a$], respectively,
we list the $b_S$ values of interest (we note that some of them are given
in Ref.\ \cite{Dover90}) below:\\
(i) For the single-channel case $\Sigma^+p\rightarrow \Sigma^+p$: 
\begin{equation}
b_0(\Sigma^+p\rightarrow \Sigma^+p)=s_1,\;\;\; 
b_1(\Sigma^+p\rightarrow \Sigma^+p)=t_2.
\end{equation}
(ii) For coupled 
channels $\Sigma^+ n$, $\Sigma^0 p$, and $\Lambda p$:
\begin{eqnarray}
b_0(\Sigma^+n\rightarrow \Sigma^+n)=\frac{1}{5}(2s_1+3s_2),\;\;\;&&
b_0(\Sigma^0p\rightarrow \Sigma^0p)=\frac{1}{10}(7s_1+3s_2),\nonumber\\
b_0(\Lambda p\rightarrow \Lambda p)=\frac{1}{10}(9s_1+s_2),\;\;\;&&
b_0(\Sigma^+n\rightarrow \Sigma^0 p)=\frac{3\sqrt{2}}{10}
(s_1-s_2),\nonumber\\
b_0(\Sigma^+n\rightarrow \Lambda p)=-\frac{\sqrt{6}}{10}
(s_1-s_2),\;\;\;&&C_0(\Sigma^0 p\rightarrow \Lambda p)=\frac{\sqrt{3}}{10}
(s_1-s_2),\nonumber\\
b_1(\Sigma^+n\rightarrow \Sigma^+n)=\frac{1}{3}(t_1+t_2+t_3),\;\;\;&&
b_1(\Sigma^0p\rightarrow \Sigma^0p)=\frac{1}{6}(t_1+4t_2+t_3),\nonumber\\
b_1(\Lambda p\rightarrow \Lambda p)=\frac{1}{2}(t_1+t_3),\;\;\;&&
b_1(\Sigma^+n\rightarrow \Sigma^0 p)=-\frac{\sqrt{2}}{6}
(t_1-2t_2+t_3),\nonumber\\
b_1(\Sigma^+n\rightarrow \Lambda p)=\frac{1}{\sqrt{6}}
(t_1-t_3),\;\;\;&&
b_1(\Sigma^0 p\rightarrow \Lambda p)=\frac{\sqrt{3}}{6}
(-t_1+t_3). \label{lpcoeff}
\end{eqnarray}
We mention that coupling between above channels has been neglected
in Ref.\ \cite{Savage96}.\\
(iii) 
Similarly there is coupling between channels containing 
$\Sigma^- p$, $\Sigma^0 n$, and $\Lambda n$, with coefficients
equal to those given in Eq.\ (\ref{lpcoeff}), except for
an overall sign change for $b_S(\Sigma^- p\rightarrow \Lambda n)$ and
$b_S(\Sigma^0 n\rightarrow \Lambda n)$, compared to
$b_S(\Sigma^+ n\rightarrow \Lambda p)$ and
$b_S(\Sigma^0 p\rightarrow \Lambda p)$.
All coefficients $b_S$ are subtraction-point dependent.

The coefficients of the $p^2$ term $C_2$ in Eq.\ (\ref{pkmin}), 
which we denote 
by $s_1',s_2',t_1',t_2',t_3'$ (apart from the common factor
$1/f^4$ to make these coefficients dimensionless), obey the same relationships
as in Eq.\ (\ref{lpcoeff}), since 
the latter are valid for the full 
momentum-dependent s-wave amplitudes \cite{Dover90}. Hence they are valid 
for the momentum expansion terms and also for tree-level amplitudes,
since the loop expansion corresponds to an expansion in powers of $\hbar$.

Let us briefly discuss 
the breaking of the SU(3)-flavor symmetry (keeping SU(2)-isospin symmetry). In a study 
of its significance it is 
concluded \cite{Dover90} that, apart from consequences of the
pseudoscalar-meson mass differences and the baryon mass differences
the SU(3) symmetry is satisfied
quite well.
The differences in baryon masses we take explicitly into account. 
The symmetry breaking
due to different meson masses is modeled by 
considering one-pion exchange (i.e.\ ignoring the kaon and the eta).
To be consistent with the above NLO approach without
mesons, we expand the exchange amplitude and keep only the leading, 
order $p^2$, term. This basically leads to addition of a symmetry-breaking term 
$-C'_2 p^2/f^4$ to tree-level amplitude, Eq.\ (\ref{treeamplitude}), 
depending on two parameters, $u_1$ and $u_2$. The relevant
coefficients in the spin-singlet channel are the following:
\begin{eqnarray}
C'_2(\Sigma^+n\rightarrow \Sigma^+n)=2 u_1,\;\;\;&&
C'_2(\Sigma^+n\rightarrow \Sigma^0 p)=-2\sqrt{2}u_1,\nonumber\\
C'_2(\Sigma^+n\rightarrow \Lambda p)=2\sqrt{2/3}u_2
,\;\;\;&&C'_2(\Sigma^0 p\rightarrow \Lambda p)=\frac{2}{\sqrt{3}}u_2.
\end{eqnarray}
The spin-triplet coefficients are obtained by multiplying the
above values with $-1/3$. 
For the channels $\Sigma^- p$, $\Sigma^0 n$, and $\Lambda n$ the
only difference is in the sign of the term
$C'_2(\Sigma^0 n\rightarrow \Lambda n)$ compared to
$C'_2(\Sigma^0 p\rightarrow \Lambda p)$.


\section{Application to hyperon-nucleon scattering data}
We now turn to the available low-energy data on hyperon-nucleon
scattering. We consider only scattering with laboratory momenta
below 200 MeV/c, i.e.\ center-of-mass momenta less than 100 MeV/c.
There is only a small amount of relevant data, corresponding to 
total cross sections for
$\Sigma^+ p\rightarrow \Sigma^+ p$, $\Lambda p\rightarrow 
\Lambda p$, $\Sigma^- p\rightarrow \Sigma^- p$, 
$\Sigma^- p\rightarrow \Sigma^0 n$ and $\Sigma^- p
\rightarrow \Lambda n$ scattering (see, for example Ref.\ 
\cite{Rijken99}). It has been known for a long time \cite{deSwart71}
that these available data do not allow a unique effective-range analysis.
The most precisely determined quantity is the ``capture ratio at rest'',
\begin{equation}
r_R=\frac{1}{4}r_0+\frac{3}{4}r_1,
\end{equation}
with
\begin{equation}
r_{S}=\frac{\sigma_{S}(\Sigma^- p\rightarrow \Sigma^0 n)}
{\sigma_{S}(\Sigma^- p\rightarrow \Sigma^0 n)+
\sigma_{S}(\Sigma^- p\rightarrow \Lambda n)},
\end{equation}
where the cross sections are taken at zero momentum and $S$
denotes the total spin.

When fitting the calculated cross sections to the data we
 make two corrections, following Ref.\ \cite{Dover90}. First, 
the 
$\Sigma^- p\rightarrow \Lambda n$ transitions are kinematically
suppressed relative to
$\Sigma^- p\rightarrow \Sigma^0 n$, as a consequence of the
large momentum release, which at threshold is 290 MeV/$c$. According
to \cite{Dover90} this necessitates inclusion of a correction factor $r_F
\approx 0$.16 in the calculated cross section for $\Sigma^- p
\rightarrow \Lambda n$.
The second correction concerns the non-negligible contribution 
of the $^3$S$_1\rightarrow ^3$D$_1$ transition in the experimentally
measured cross section $\sigma_1(\Sigma^- p\rightarrow \Lambda n)$.
Defining the ratio $r_{ds}$ of $^3$S$_1\rightarrow ^3$D$_1$ to
$^3$S$_1\rightarrow ^3$S$_1$ cross sections for $\Sigma^- 
p\rightarrow \Lambda n$, a value of $r_{ds}\approx 0.57$ was
estimated in Ref.\ \cite{Dover90}.

We first attempted a fit of 28 data points (the cross sections shown in
Fig.\ 1. and the capture ratio $r_R=0.468\pm 0.010$) with the leading-order 
calculation involving the 5 
parameters $s_1,s_2,t_1,t_2,t_3$, 
varying the value of the subtraction point $\mu$. In this leading-order 
calculation we could not achieve a reasonable fit.

Then we turned to the next-to-leading-order computation, 
which added the 5 SU(3)-symmetry
respecting parameters $s'_1,s'_2,t'_1,t'_2,t'_3$, as well as the 2 symmetry
breaking ones, $u_1$ and $u_2$.
As further constraint we used 
the existence of the hypertriton bound 
state \cite{Miyagawa95}, which is compatible with the OBE  
potential NSC97f of Ref.\ \cite{Rijken99}. 
The advantage of this \cite{deSwart71} is that while the scattering
data are mainly sensitive to the $^3$S$_1$ $\Lambda N$ interaction,
the hypertriton is more sensitive to the $^1$S$_0$ $\Lambda N$ interaction.
Thus we imposed as 
requirement on the fit that it had to lead to $\Lambda$-nucleon
scattering lengths of $a_{\Lambda N}^{(0)}\approx -2.5$ fm for the
singlet, and $a_{\Lambda N}^{(1)}\approx -1.75$ fm for the triplet 
state, respectively.

Several
fits of the same quality, $\chi^2$ being around 13 (for 16 degrees of
freedom) could be obtained. If a fit was obtained for a certain value of the 
subtraction
point $\mu$, it is possible to vary $\mu$ in the range of
0.1 GeV to 0.3 GeV, leading to practically unchanged scattering 
amplitudes and thus the same medium effects. We do not consider members 
of such a family of fits as different. The slight change
in $\chi^2$ corresponds to variation (in a certain range) of 
the parameters of terms proportional 
to $p^2$ in the tree amplitudes, thus changing the effective-range
values. This is not surprising 
considering the rather large error bars on the measured cross sections. 
As a
consequence, the parameters $s'_1, s'_2,t'_1,t'_2,t'_3, u_1,u_2$ are not very
well determined allowing fits with different effective-range values.
Our experience shows that all these fits lead to 
quite similar scattering amplitudes and predictions for in-medium mass shifts.

Our strategy in dealing with these fits is to accept only those which
do not imply spurious (quasi-)bound states (``hyperdeuterons'') near threshold.
For example, a fit with the lowest $\chi^2$ led to negative effective range 
in the spin-singlet $\Sigma^+ p$ channel, $r^{(0)}_{\Sigma^+ p}=
-2.32$ fm. Together with the corresponding scattering length 
$a^{(0)}_{\Sigma^+ p}=
2.64$ fm it would imply \cite{Lutz00} a bound state
with binding energy of 3.3 MeV, which is known not to exist. Therefore
we rejected this fit.

With $\chi^2 = 13.5$ we obtained a fit with reasonable values of
the range parameters and $\Lambda N$ scattering lengths close to 
values of model NSC97f in Ref.\ \cite{Rijken99}. The  
corresponding set (denoted by A) of parameters 
is the following: $\mu=0.15$ GeV, $s_1=
  1.00, s_2=    
   1.78, t_1=    
   1.21, t_2=    
   3.43, t_3=    
   1.87, s'_1=    
   -0.447, s'_2=    
  1.82, t'_1=    
   -2.42, t'_2=    
  1.05, t'_3=    
  3.60, u_1=    
   -0.260,$ and $u_2=
   -0.0998$. The resulting cross sections are shown in Fig.\ 1.
The calculated value of the capture ratio is
$r_R=0.4670$, the $\Lambda N$ scattering lengths and 
range parameters are:
$a_{\Lambda N}^{(0)}=-2.50$ fm, $r_{\Lambda N}^{(0)}=1.61$ fm,
$a_{\Lambda N}^{(1)}=-1.78$ fm, $r_{\Lambda N}^{(1)}=1.42$ fm.
For $\Sigma^+ p\rightarrow \Sigma^+ p$ scattering we obtain then:
$a_{\Sigma^+ p}^{(0)}=0.55$ fm, $r_{\Sigma^+ p}^{(0)}=0.36$ fm,
$a_{\Sigma^+ p}^{(1)}=0.94$ fm, $r_{\Sigma^+ p}^{(1)}=0.35$ fm.
These values are quite different from those of model NSC97f.

\vspace*{10mm}
\epsfxsize=14cm
\centerline{\epsffile{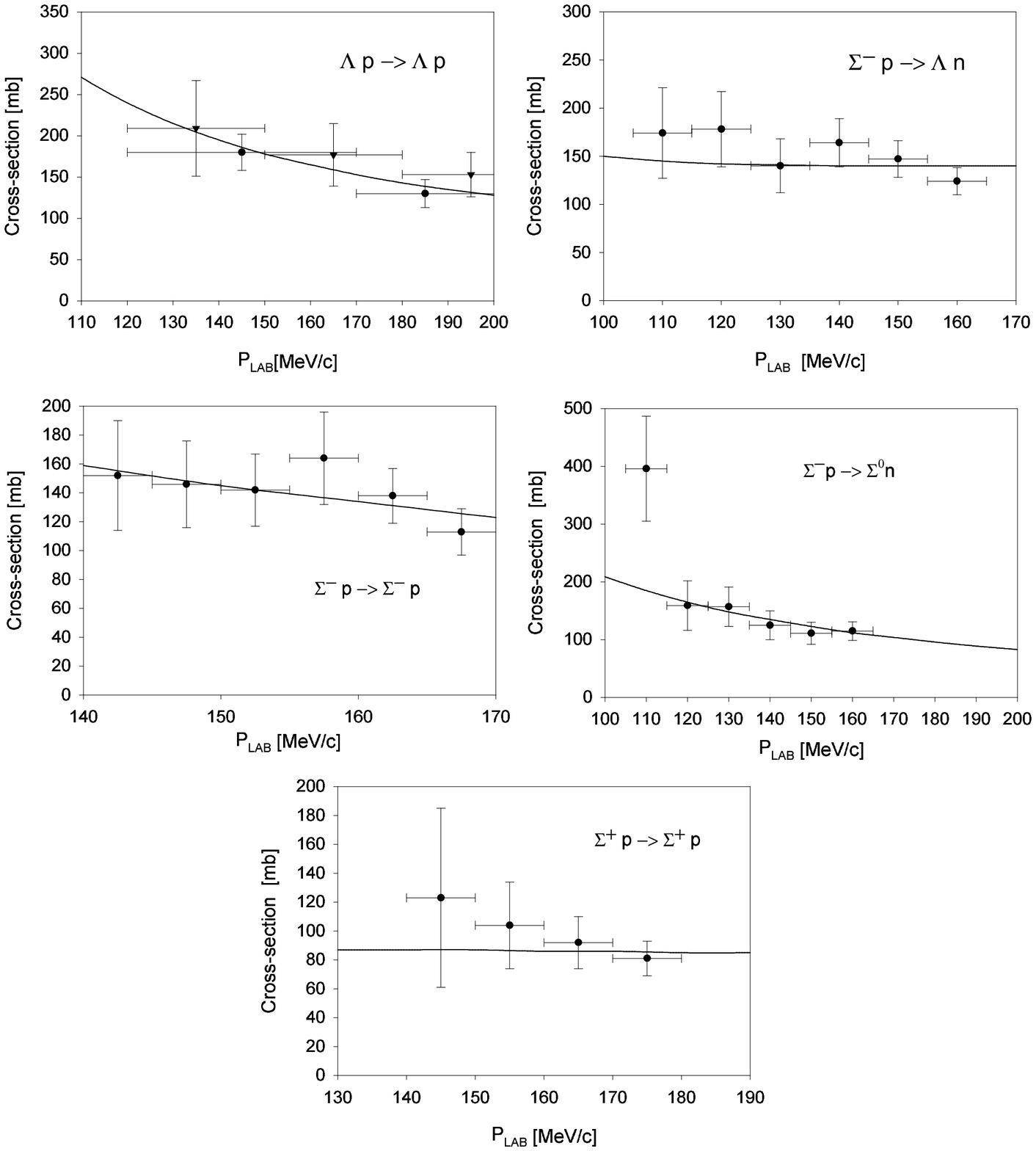}}
\begin{center}
Fig.\ 1: 
Results of the fit (line) to $YN$ cross sections with parameter set A.
\end{center}

Our results high-light one of the main problems
with the available hyperon-nucleon scattering data (apart from 
the large error bars): the absence of truly low-energy cross sections.
Fig.\ 1 shows that the lowest hyperon (laboratory) momentum is larger
than 100 MeV/$c$, which means that the reciprocal of the scattering 
length and the range term, $\frac{1}{2}rp^2$, can be of the same order.
This leads to results for the scattering lengths and effective ranges 
that are not unique.

We mention that we could not obtain a satisfactory fit to the data with 
both $\Lambda$-nucleon
and $\Sigma^+$-proton scattering lengths being close to the values given 
in Ref.\ \cite{Rijken99}.

\section{Hyperon mass shifts in the nuclear medium}
Since we have dealt only with the low-momentum expansion of the hyperon-nucleon
interaction, we can only consider hyperon self-energies in the
nuclear medium in a low-density expansion \cite{Dover71}. A simple
expression relates the in-medium mass shift (i.e. the self-energy at zero
momentum) to the hyperon-nucleon scattering lengths \cite{Savage96}, viz.
\begin{equation}
\Delta M_Y=\frac{2\pi}{M_r}
\left[
\left(\frac{a_{Yn}^{(0)}}{4}+\frac{3a_{Yn}^{(1)}}{4}\right)\rho_n
+
\left(\frac{a_{Yp}^{(0)}}{4}+\frac{3a_{Yp}^{(1)}}{4}\right)\rho_p
\right],
\end{equation}
where $M_r$ is the reduced mass and $\rho_n$ ($\rho_p$) is the 
neutron (proton) density.

A better approximation is obtained when one takes into account
the Fermi motion of the nucleons, which amounts to integrating
the momentum-dependent forward-scattering amplitude in the
nucleon Fermi sea. In this case one has
\begin{eqnarray}
\Delta M_Y&=&-\frac{1}{4}
\left[
2\int_0^{p_{F_n}} \left({\cal A}_{Yn}^{(0)}(0,\vec p;0,\vec p)+3
{\cal A}_{Yn}^{(1)}(0,\vec p;0,\vec p)\right)\frac{d^3 p}{(2\pi)^3}
\right.
\nonumber\\
&&+\left.
2\int_0^{p_{F_p}} \left({\cal A}_{Yp}^{(0)}(0,\vec p;0,\vec p)+3
{\cal A}_{Yp}^{(1)}(0,\vec p;0,\vec p)\right)\frac{d^3 p}{(2\pi)^3}
\right],\label{ampint}
\end{eqnarray}
with ${\cal A}^{(0)}$ (${\cal A}^{(1)}$) being the singlet (triplet)
amplitude, $\vec p$ is the nucleon momentum and that of the hyperon 
is zero.

Since our fits to cross sections extend only up to momenta of 200 MeV/$c$,
that leads 
to a maximum density of 0.4$\rho_0$ for
the case of isospin-symmetric matter and 0.2$\rho_0$ for neutron
matter ($\rho_0=0.17\;\rm{fm}^{-3}$ is the saturation density).
Results for hyperon mass shifts for the parameter set A are shown
in Fig.\ 2 for isospin-symmetric matter (a), and pure neutron
matter (b).

\vspace*{10mm}
\epsfxsize=14cm
\centerline{\epsffile{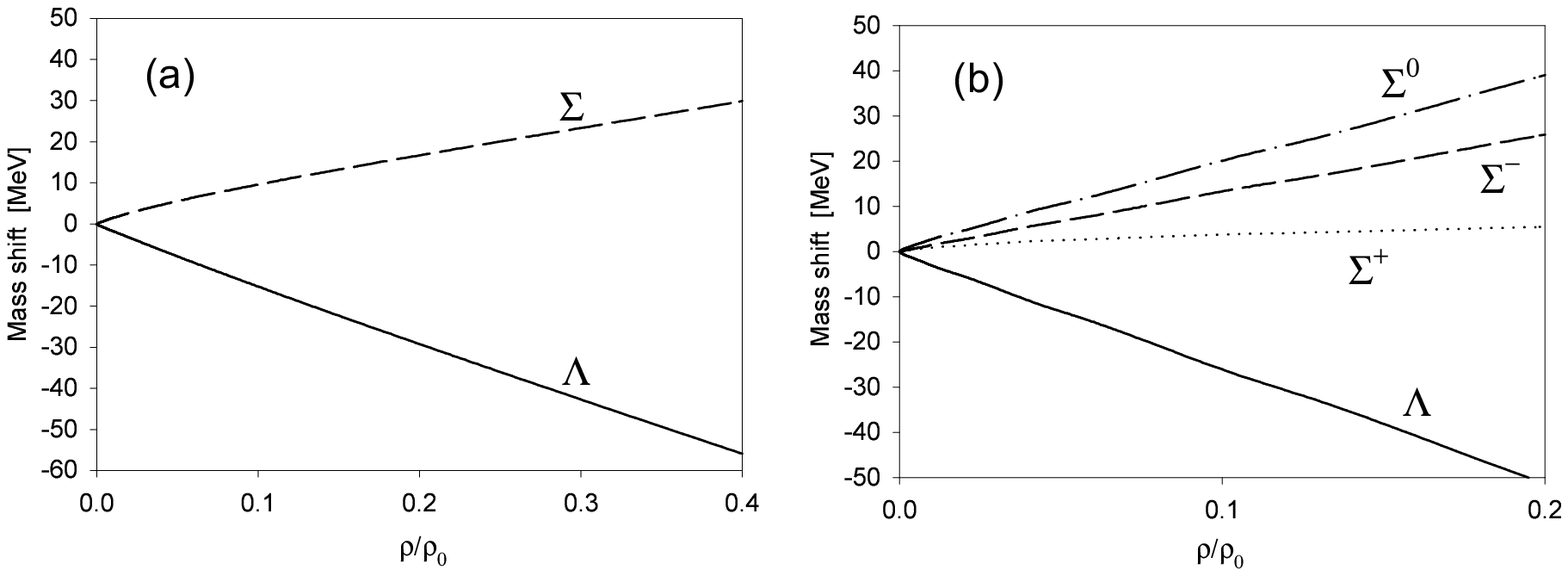}}
\begin{center}
Fig.\ 2:
Hyperon mass-shifts in isospin-symmetric matter (a) and
neutron matter (b), calculated with parameter set A.
\end{center}

The results for the $\Lambda$ mass shift in Fig.\ 2(a) are similar to those
obtained in Refs.\ \cite{Schulze98,Schulze95,Baldo98,Schulze01} using the 
Brueckner-Hartree-Fock
(BHF) method with the OBE potential from Ref.\ \cite{Maessen89}, 
except that there the linear behavior holds up to 0.2$\rho_0$,
after which nonlinear behavior, leading eventually to saturation, sets in.
For the $\Sigma$ mass shift the results differ, since in Ref.\ \cite{Schulze98} a small 
mass decrease is found at small nucleon density. 
We mention that the hypertriton calculation 
\cite{Miyagawa95} shows that the hyperon in $^3_\Lambda$H is a $\Lambda$ with 
probability greater than 95\%, thus the behavior of the $\Sigma$ 
in nuclear medium has a very small effect on the binding energy.

Results for neutron matter cannot be compared directly with those
of the BHF method (and potentials used), since only larger densities
were considered in Refs.\ \cite{Schulze98,Schulze95,Baldo98}, as well as in 
Ref.\ \cite{Yamamoto00}. Low-density results \cite{Motoba} show linear
behavior until about $0.2 \rho_0$ and striking differences for different
potentials.  
Mass shifts based on parameter set A, Fig.\ 2(b), show similar
behavior to these BHF results 
\cite{Schulze98,Schulze95,Baldo98,Yamamoto00},
with the difference that the latter predict the $\Sigma^0$ curve
below that of $\Sigma^-$. We mention that in this case, because of
isospin-symmetry violation in the neutron matter, the nondiagonal mass shift
$\Delta M_{\Lambda\Sigma}$ is nonzero and quite large for our
parameter values, leading to strong repulsion of the 
$\Lambda$ and $\Sigma^0$ curves. Without this repulsion we would recover
the ``natural'' ordering (in increasing mass shift) of the $\Sigma^+, 
\Sigma^0, \Sigma^-$ curves. Also, the BHF calculations seem to 
prefer a mass decrease for the $\Sigma$-hyperons (except for $\Sigma^-$
and, depending on the potential, also for $\Sigma^0$), compared to 
our repulsion.

We conclude that due to the scarce scattering data the predicted hyperon
mass shifts in the nuclear medium are somewhat uncertain, but the
uncertainty can be partly eliminated using conclusions from the existence of
the hypertriton bound state and imposing a constraint of reasonable 
effective-range values, thus avoiding spurious hyperon-nucleon bound states. 
Computations in the BHF scheme
provide more reliable results for larger densities, but they 
suffer from serious ambiguities stemming from the use of different potentials, all
of which describe the scattering data equally well.

\acknowledgments
We acknowledge useful discussions with U.\ van Kolck
and T.\ Motoba, as well as helpful
correspondence with H.-J.\ Schulze and Y.\ Yamamoto.
This research was supported in part by the
Hungarian Research Foundation (OTKA) grant T030855
and by the NWO-OTKA grant 834012.
The research of RGET was made possible by a fellowship
of the Royal Netherlands Academy of Arts and Sciences. That
of AELD is part of the research program of the 
``Stichting voor Fundamenteel Onderzoek der Materie" (FOM) with
financial support from the
``Nederlandse Organisatie voor Wetenschappelijk Onderzoek" (NWO).


\end{document}